\documentclass[12pt]{article}

\usepackage{amsmath}
\usepackage{graphicx}
\usepackage{amssymb}
\usepackage{amsfonts}
\usepackage{latexsym}
\usepackage{color,soul} 

\def\beq{\begin{eqnarray}}
\def\eeq{\end{eqnarray}}

\def\al{\alpha}
\def\be{\beta}

\def\ka{\kappa}

\def\si{\sigma}

\def\La{\Lambda}

\begin{document}

\begin{center}

{\large\bf On the gravitational seesaw in higher-derivative gravity}
\vskip 8mm

Antonio Accioly$^{a}$,
\ \
Breno L. Giacchini$^{a}$,
\ \
Ilya L. Shapiro$^{b,c}$

\end{center}
\vskip 1mm

\begin{center}
{\sl
(a) Centro Brasileiro de Pesquisas F\'{\i}sicas
\\
Rua Dr. Xavier Sigaud 150, Urca, 22290-180, Rio de Janeiro, RJ, Brazil
\vskip 3mm

(b) \ Departamento de F\'{\i}sica, ICE, Universidade Federal de Juiz de Fora
\\
Campus Universit\'{a}rio - Juiz de Fora, 36036-330, MG, Brazil
\vskip 3mm

(c) \ Tomsk State Pedagogical University and Tomsk State
University, Tomsk, Russia
}
\vskip 2mm\vskip 2mm

{\sl E-mails:
\ \
accioly@cbpf.br,
 \
breno@cbpf.br,
 \
shapiro@fisica.ufjf.br}

\end{center}
\vskip 6mm

\begin{quotation}
\noindent
\textbf{Abstract.} \
Local gravitational theories with more than four derivatives are
superrenormalizable, and also may be unitary in the Lee-Wick
sense. Thus, it is relevant to study the low-energy  properties of
these theories, especially to identify observables which might be
useful for experimental detection of higher  derivatives. Using
an analogy with the neutrino Physics, we explore the possibility
of a gravitational seesaw mechanism, in which several dimensional
parameters of the same order of magnitude produce a hierarchy
in the masses of propagating particles. Such a mechanism could
make a relatively light degree of freedom detectable in low-energy
laboratory and astrophysical observations, such as torsion balance
experiments and the bending of light. We demonstrate that such a
seesaw mechanism in the six- and more-derivative theories is
unable to reduce the lightest mass more than in the simplest
four-derivative model. Adding more derivatives to the
four-derivative action of gravity makes heavier masses even
greater, while the lightest massive ghost is not strongly affected.
This fact is favorable for protecting the theory from instabilities,
but makes the experimental detection of higher derivatives more
difficult.
\vskip 3mm

\vskip 2mm

\vskip 2mm

\end{quotation}

\section{Introduction}

The role of higher derivatives in quantum and classical gravity
theories is important, complicated and ambiguous. On the one hand
it is well-known that semiclassical~\cite{UtDW} and quantum
\cite{Stelle77} gravity can be formulated as renormalizable
theories only with the four-derivative terms in the action
(see~\cite{birdav,book} for an introduction and~\cite{PoImpo}
for a recent review). On  the other  hand, by adding higher-derivative
terms to the Einstein-Hilbert action one introduces massive unphysical
ghosts, related instabilities and (in the quantum gravity case) a
non-unitary $S$-matrix. Recently, it was shown that in a theory
with six or more derivatives one can have all massive poles
complex, and then the $S$-matrix becomes unitary in the Lee-Wick
sense~\cite{LQG-D4}.

Let us remember that higher derivatives emerge also in the
gravitational effective action in string theory. The corresponding
terms are removed by means of the Zweibach reparametrization of
the background metric in target space~\cite{zwei}. However, this
procedure is ambiguous, since the no-ghost condition does not fix
many terms in the higher derivative sector~\cite{maroto}.
Furthermore, another source of ambiguity is that the problem
may be solved not only by completely removing all potentially
dangerous terms, but also by reducing the effective action to
a ghost-free non-local form~\cite{Tseytlin-95}.

It is important  to note that in both these approaches the removal
of  massive ghosts requires an {\it absolutely precise} fine-tuning
of the action. Nevertheless, any small violation here should lead to
destructive instabilities and, moreover, these instabilities are even
stronger for smaller violations~\cite{Woodard-r}. This means that
the ghost-killing procedure in string theory~\cite{zwei} (or
\cite{Tseytlin-95}) demands an  absolutely precise fine-tuning
of infinitely many parameters. On the other hand, violations of
the fine tuning can not be avoided if the loop contributions are
taken into account~\cite{CountGhosts} in the effective  field theory
framework (see, e.g.,~\cite{Burgess}).  The most reasonable position
is all in all to assume the existence of higher derivatives and
try to understand why they do not produce a total destruction
of the classical gravitational
solutions~\cite{GW-Stab}\footnote{ The results of this work are
coherent with the previous works on stability of de Sitter space in
fourth order gravity, which was first considered in \cite{MSS}.
A more detailed analysis of cosmic perturbations in four-derivative
gravity, with qualitatively similar conclusions, was given in a recent
work \cite{IT}.}.

Keeping the string theory in mind, one can assume that the
action of the theory has only one fundamental dimensional
parameter, that is the Planck mass. All dimensionless coefficients
are supposed to be of order one. Hence all phenomena which occur
at sub-Planckian energies may be considered as low-energy ones.
Then, assuming that there is no fine-tuning and that the higher
derivative terms are there, the natural questions are: \ {\it i)}~do
we have a chance to see the effect of higher-derivative terms at
low energies?  \ {\it ii)}~is the IR gravitational physics protected
from the ghosts, if the dimensional parameters are all related
to the Planck mass? These questions are particularly relevant,
because already at the semiclassical level the loop corrections
produce non-local form factors in the quadratic curvature terms.

At low energies, for the sake of simplicity it is
natural to assume a truncation of the infinite series in the
d'Alembert operator, leading  to an effective polynomial theory
of the type~\cite{highderi}
\beq
S
&=& \frac{1}{16\pi G}
\int d^4x\sqrt{-g}\,\big(R-2\La \big)
\nonumber
\\
& & +\,\int d^4x\sqrt{-g}\,\Big\{
c_1R_{\mu\nu\al\be}^2 + c_2R_{\mu\nu}^2 + c_3R^2
\nonumber
\\
& &+ \,
d_1 R_{\mu\nu\al\be} \Box R^{\mu\nu\al\be}
+ d_2 R_{\mu\nu}\Box R^{\mu\nu}
+ d_3 R\Box R
\nonumber
\\
& & + \,d_4R^3 \, + \,d_5 R R^{\mu\nu}R_{\mu\nu} \,+ \dots + f_1 R_{\mu\nu\al\be} \Box^k R^{\mu\nu\al\be}
\nonumber
\\
& & + f_2 R_{\mu\nu}\Box^k R^{\mu\nu}
+ f_3 R\Box^k R\,+ \dots +\, f_{...}R_{...}^{k+2}
\Big\}\,,
\label{superre}
\eeq
where we have used the same sign conventions as in~\cite{Accioly15}.

In what follows we will be interested in the modified Newtonian
potential and the bending of light by
a weak gravitational field. In this spirit, we can disregard the
cosmological constant term and those terms which are third-
or higher-order in curvature. Furthermore, for the sake of simplicity
we start the analysis from the $k=1$ case. As a consequence, the
relevant part of the action can be cast into the form
\beq
S
&=&
S_{grav}\,+\,\int d^4 x \sqrt{-g} \,{\cal L}_m\,,
\label{totaction}
\\
S_{grav}
&=&
\int d^4 x \sqrt{-g} \Big\{
\dfrac{2}{\kappa^2} R + \dfrac{\alpha}{2} R^2
+ \dfrac{\beta}{2} R_{\mu\nu}^2
 + \dfrac{A}{2} R \square R
+ \dfrac{B}{2} R_{\mu\nu} \square R^{\mu\nu} \Big\} \,,
\label{Lag6orderGravity}
\eeq
where an additional matter action was introduced; besides,  the
notations were adjusted for the sake of consequent calculations.
Here $\,\al$, $\be$, $A$ and $B$ are free parameters, where the
first two are dimensionless while $A$ and $B$ carry dimension
of (mass)$^{-2}$. In what follows we will refer to the quantities
$|B|^{-1/2}$ and $|A|^{-1/2}$ as to the massive parameters of
the action. The notation $\kappa^2/2=16\pi G = M_P^{-2}$ is
conventional in the quantum gravity literature.

As we have mentioned above, in string theory all massive parameters
are constructed from the single dimensional parameter $\al^\prime$,
and hence all masses in the action are supposed to have the same
(typically Planck) order of magnitude. However, our experience with
the seesaw mechanism in neutrino Physics shows that this does not
rule out a situation where several huge massive parameters combine
into one particle  of light mass, with the other masses becoming even
greater. In our case the quantities in the action must satisfy
$A^{-1},\,B^{-1},\,\ka^{-2} \sim M_P^2$. In what follows we
discuss the possibility of a  seesaw-like mechanism. As we shall
see, in the gravitational case it enables one to have a parameter
$B^{-1}$ much smaller than $M_P^2$ and still have an associate
mass of the order of $M_P$. This scenario can be achieved by
reducing the lighter mass of the tensor excitation, which is the
well-known ghost mode. 

The paper is organized as follows. In Sec.~\ref{S2} we discuss the
new gravitational seesaw mechanism, in the theory with more than
four derivatives, such as in~(\ref{Lag6orderGravity}).
In Sec.~\ref{Scomplex} the results of the previous section are
extended to the case in which the propagator has complex poles.
Possible observational effects caused by a light particle are
briefly commented in Sec.~\ref{S3}. We remark that the main
focus of this communication is on the seesaw mechanism, while
the detailed discussion concerning phenomenological aspects of
the theory with six or more derivatives of the metric will be
given in the parallel work~\cite{Larger}, devoted to the modified
Newtonian potential and the bending of light. Finally, in
Sec.~\ref{S5} we draw our conclusions.

\section{Gravitational seesaw in higher derivative theories}
\label{S2}

The conventional point of view is that  higher derivatives are not
observable at low energies because of the Planck suppression. In
order to have the Planck suppression in four-derivative gravity, the
coefficients of the higher-order terms have to be of order one or
at least not too many orders of magnitude greater. However, what
is correct as far as  the four-derivative model is concerned, is not
necessary right for theories exhibiting  six derivatives or more.
Since there are several massive parameters, one can imagine a
specific seesaw-like mechanism, which enables two (or more)
large-mass parameters to combine in such a way that they
produce a much smaller physical mass. Let us examine
the theory~(\ref{Lag6orderGravity}) in this respect.

In the weak-field limit, i.e.
$g_{\mu\nu} = \eta_{\mu\nu} + \kappa h _{\mu\nu}$ and
$\vert \kappa h _{\mu\nu} \vert \ll 1$, the linearized field
equations can be cast into the form
\beq
\label{eqMotion6LinA}
&&
\left( \frac{2}{\kappa^2}
- \frac{\beta}{2} \square - \frac{B}{2} \square^2\right)
\left( R_{\mu\nu} - \frac{1}{2} \eta_{\mu\nu} R \right)
\nonumber
\\
&-& \left( \alpha + \frac{\beta}{2} + A \square
+ \frac{B}{2} \square \right) \left( \eta_{\mu\nu} \square R
- \partial_\mu \partial_\nu R \right) = - \frac{T_{\mu\nu}}{2}\,.
\nonumber
\\
\eeq
It is possible to show that, introducing a suitable gauge condition,
the weak gravitational field generated by a static point-like mass,
$\,T_{\mu\nu}(\textbf{r})
 = M \eta_{\mu 0} \eta_{\nu 0} \delta^{(3)}(\textbf{r})$,
has non-zero components given by (one can find
more detailed and general results in \cite{Newton-high,Giacchini-poles})
\beq
\label{solutionFinal}
h_{00}
&=&
\frac{M\kappa}{16\pi} \Big( - \frac{1}{r}
+ \frac{4}{3} F_2 - \frac{1}{3}F_0\Big)\,,
\nonumber
\\
h_{11}
&=&
 h_{22} \,=\, h_{33} \,=\, \frac{M\kappa}{16\pi}
 \Big( - \frac{1}{r} + \frac{2}{3}\,F_2 +\frac{1}{3}\,F_0\Big)\,,
\eeq
where
\beq
F_k =
\frac{\mu_{k+}^2}{\mu_{k+}^2 - \mu_{k-}^2} \frac{e^{-\mu_{k-}r}}{r}
+ \frac{\mu_{k-}^2}{\mu_{k-}^2 - \mu_{k+}^2} \frac{e^{-\mu_{k+}r}}{r}\,.
\nonumber
\eeq
Here $\,k=0,2\,$ labels the spin of the particles, whose masses are
defined by the positions of the poles of the propagator,
\beq
\label{Def_masses}
\mu_{2\pm}^2
= \dfrac{\beta \pm \sqrt{\be^2 + \frac{16}{\ka^2}B}}{2 B},
\quad
\mu_{0\pm}^2 =
\dfrac{\si_1 \pm \sqrt{\si_1^2
- \frac{8\si_2}{\ka^2}}}{2 \si_2},
\nonumber
\\
\eeq
with $\sigma_1 \equiv 3\alpha + \beta$ and $\sigma_2 \equiv 3A+B$.
One can observe that in the sixth-order gravity massive particles
occur in dependent pairs with the same spin.
The masses (\ref{Def_masses}) are real and non-degenerate
provided that
\beq
&&
\be,\,B < 0,
\quad
\sigma_1,\sigma_2>0
\quad
\mbox{and}
\nonumber
\\
&&
\beta^2  + \frac{16B}{\kappa^2} > 0\,,
\quad
\mbox{while}
\quad
\si_1^2 - \frac{8\si_2}{\ka^2}> 0\,.
\label{signs}
\eeq
Indeed, the quantities $\mu_{0\pm}$ and $\mu_{2\pm}$
could be complex and still yield a real solution to the equations
of motion (\ref{eqMotion6LinA}) and thus physically admissible
results, e.g., through a real effective
potential~\cite{Larger,Giacchini-poles}.
Here, however, we restrict the analysis to the case of real poles,
while the scenario with complex poles is explored in the
following section.

Let us start the discussion of the mass relations in the
six-derivative theory from considering  the tensor sector.
According to  Eq.~(\ref{Def_masses}), for the case of real poles
with  $\mu_{2+}^2 < \mu_{2-}^2 \,$, the lighter massive excitation
is a ghost and the other is a healthy tensor field~\cite{highderi}.
Using the Eqs.~(\ref{Def_masses}) and (\ref{signs}) it is easy to
show the existence of a relation between $\be$ and $B$, namely
\ $16 \vert B \vert \ll \ka^2 \be^2$,  in the special case when one
of the masses is much smaller than the other,
\beq
\mu_{2+}^2 \ll \mu_{2-}^2 \,.
\label{seesaw}
\eeq
In the theory where this condition is satisfied, the masses
$\mu_{2\pm}$ can be approximated by
\beq
\mu_{2+}^2 \approx \frac{4}{\ka^2 \vert \be \vert}
\,\,\ll\,\,
\mu_{2-}^2 \approx \frac{\be}{B}.
\label{8}
\eeq

As in the original neutrino's seesaw mechanism one of the masses
depends, roughly, on only one parameter, while the other depends
on both. Moreover, this relation occurs in such a manner that if
the lighter mass is reduced, then the larger mass is augmented.
A remarkable difference with respect to the neutrino's
mechanism is that while in the neutrino case it works to make the
lightest mass even lighter, in the gravitational model the effect
is to turn the largest mass even larger, according to Eq.~(\ref{8}).
This happens due to the presence of the parameter
$B$ in the denominator of Eq.~(\ref{Def_masses}), making the
lightest mass to depend only on $\be$ while the largest one
depends on both parameters.

In this vein, there are two possible ways of having $\mu_{2-}$ of
the order of the Planck mass: \ to have a small $|B|$ or a larger
$|\be|$. The first choice is the standard one, since it prescribes
that $\be \sim 1$ and $B \sim M_P^{-2}$ so as to have all the masses
to the order of $M_P$. The second possibility, which relies on the
seesaw mechanism, allows one to have $|B| \gg M_P^{-2}$ and
still have $\mu_{2-} \sim M_P$. Of course, having a large $|B|$ still
yielding one large mass can only be achieved by means of the ghost
mass reduction trough a parameter $\be \gg 1$. The final result,
which can be seen from Eq.~(\ref{8}), is that the existence of a much
lighter mass of the first (ghost) state depends only on the
second- and fourth-derivative terms, while the six-derivative
term does not affect the presence of much lighter mass.
\textit{Mutatis mutandis} these arguments also apply to the
scalar modes. By the end of  the day, the six-derivative terms are
not capable to produce an efficient seesaw mechanism working
like in the case of the neutrino mass.

One can present a general argument in favour of the non-possibility
of the strong seesaw mechanism for even higher-order, i.e. eight
and more, gravity theories. For instance, consider the action
(\ref{superre}) with $k=2$, that means eight-derivative theory.
One can write the equation for the massive poles in the propagator
in the form
\beq
\frac{1}{m_0^4}\,k^6 \,-\,
\frac{3}{m_1^2}\,k^4 \,+\,
3\be\,k^2 \,-\,
m_2^2 \,=\,0\,.
\label{8der pole}
\eeq
Here $m_{0,1,2} \,$ are positive massive parameters coming from
the action. In string theory one can assume that they are all of
the same order of magnitude, say
\beq
m_0^2 \,\sim\, m^2_1 \,\sim\,  m^2_2 \,\sim\,  M_P^2\,.
\label{sim}
\eeq
Let us assume that this is the case. One can rewrite (\ref{8der pole})
in the more simple form
\beq
k^6 \,-\,
\frac{3m_0^4}{m_1^2}\,k^4 \,+\,
3\be\,m_0^4\, k^2 \,-\,
m_0^4m_2^2 \,=\,0\,.
\label{8der}
\eeq
The roots of this equation are defined by the Cardano formula and
can be either real or complex. Consider the particular case of
real positive roots which satisfy the hierarchy
$\mu^2_1 \ll \mu_2^2 \sim \mu^2_3$. Then the equation
becomes
\beq
k^6 &-&
\big(\mu^2_1 +\mu_2^2 + \mu^2_3\big)\,k^4
+ \, \big(\mu^2_1\mu_2^2
+ \mu^2_1\mu^2_3+ \mu^2_2\mu^2_3\big)\, k^2
-
\mu^2_1\mu^2_2\mu^2_3 \,=\,0\,.
\label{8der rooted}
\eeq
Using the hierarchy $\mu^2_1 \ll \mu_2^2 \sim \mu^2_3$,
the last equation boils down to
\beq
k^6 \,-\,
\big(\mu_2^2 + \mu^2_3\big)\,k^4
\,+\,
\mu^2_2\mu^2_3\, k^2 \,-\,
\mu^2_1\mu^2_2\mu^2_3 \,=\,0\,.
\label{8der r}
\eeq
It is easy to see that there is a contradiction between
Eq. (\ref{8der}) with (\ref{sim}) and Eq. (\ref{8der r}).
According to (\ref{8der}) we have
\beq
&&
\frac{3m_0^4}{m_1^2}\,\sim\,M_P^2\,,\quad
3\be\,m_0^4\,\sim\,M_P^4
\qquad
\mbox{and}
\qquad
m_0^4m_2^2 \,\sim\,M_P^6\,.
\label{contra}
\eeq
However, this does not fit  Eq.~(\ref{8der r}), because the
last requires
\beq
&&
\mu_2^2 + \mu^2_3 \sim M_P^2
\,,\quad
\mu^2_2\mu^2_3 \sim M_P^4
\qquad
\mbox{ but } \qquad
\mu^2_1\mu^2_2\mu^2_3 \ll M_P^6\,.
\label{8der contra}
\eeq
This consideration can be easily extended to the higher number
of derivatives, and the result will be always the same. We leave it
as an exercise to the interested reader. It is also worth stressing
that this general reasoning applies to both tensor and scalar
sectors of the model. Finally, the mechanism that actually may
take place in higher-derivative gravity can be called ``weak
seesaw''. A larger mass can become even larger, while a smaller
one does not become smaller without using unnatural values
for the dimensionless parameters of the action.

The main conclusion is that the real  poles of the propagator can
not provide a much smaller mass of the lightest ghost constructed
from the coefficients which are all of the Planck order of magnitude.
Is it bad or not, from the Physics side? We know that the presence
of ghost means potential instability, but in the case of gravity the
situation may be different \cite{GW-Stab}, for instance because
of the singular nature of non-polynomial theory which escapes
the Ostrogradsky  instability \cite{Woodard-r}. Since a consistent
theory of quantum or semiclassical gravity without higher
derivatives looks impossible,  the general situation with  stability
looks unclear and it makes sense to assume that ghosts exist but
for some reason they do not lead to a fast decay of the vacuum
and other type of instabilities. The existing explanation for this
is related to the huge mass of the ghost \cite{GW-Stab}
(not tachyon! --- see~\cite{GiuFil}) which does not permit the
creation of a ghost particle from vacuum without generating
Planck-order density of gravitons. From this perspective it is
important that the mass of the lightest ghost is protected from
the seesaw mechanism if even more derivatives are added to
the action~(\ref{superre}).

The mass of the lightest scalar (healthy) excitation is likewise
protected from the seesaw mechanism. Let us note that the
successful realization of the Starobinsky inflation model
\cite{star} requires a large value of the coefficient of the
$R^2$ term \cite{star83,KS-2012}. Such a coefficient is
reducing the mass of the scalar mode considerably, but this
happens without strong seesaw mechanism. It remains to see
what would be further phenomenological consequences of the
light ghost (and/or of a light scalar particle) within the weak
seesaw mechanism as in (\ref{seesaw}). In Sec.~\ref{S3} we
consider an example of this kind.

\section{Seesaw with complex poles}
\label{Scomplex}

In the previous section we have dealt only with the case in which
the propagator has real poles. Of course, if the quantities $\mu_{i}$
are complex it does not make sense to consider a seesaw-like mechanism
meaning a strong hierarchy between those ``masses''. This idea, though, can
be extended keeping in mind the original motivation for considering
a gravitational seesaw: to have huge-mass parameters in the action
resulting in a small physical particle mass. Hence, we shall define
the seesaw mechanism in the case of complex poles as a way of having
huge-mass parameters in the action yielding small physical massive
parameters, which turn out to be the real and the imaginary parts of
$\mu_{i}$.

Instead of starting from the six-derivative gravity example as we
proceeded for the case of real poles, now we shall go directly to the
general proof outlined in the previous section. Since the poles of the
propagator are defined as the roots of
a polynomial equation such as~\eqref{8der pole}, it follows from the fundamental
theorem of algebra that complex poles always occur in
conjugate pairs. Let $\, \mu_1 = a + i b \,$ and
$\, \mu_{1^\prime} = a - i b \,$ be one such pair. Then, writing
the equation for the poles
in terms of the roots, as in~\eqref{8der rooted}, the coefficient
formed by the sum of all the roots (squared) will contain
the term $2(a^2 - b^2)$. The coefficient which involves the products of the
roots chosen two by two will have the terms
\beq
\mu_1^2 \mu_{1^\prime}^2 &=& a^4 + b^4 + 2 a^2 b^2,
\label{product2}
\\
\mu_1^2 \mu_2^2 + \mu_{1^\prime}^2 \mu_2^2 &=& 2 (a^2 - b^2) \mu_2^2 ,
\eeq
for
an arbitrary third root $\mu_2^2$, and so on.

Of course, all the coefficients
will be real, since the parameters of the action are also real. In this
sense there is not much difference with the case of real poles, apart
from the fact that now the relevant quantities are $a^2$ and $b^2$. The
last term, however, formed by the product of all the roots, will contain
the term $\mu_1^2 \mu_{1^\prime}^2$ given by~\eqref{product2}. This
conversion of the product of all the roots into a sum changes the argument
used in the last section to show that the strong seesaw mechanism does not
work. In fact, thinking on the eight-derivative model above, it is well
possible to have $m_0^4m_2^2 \sim M_P^6$ in Eq.~\eqref{contra} at the
same time in Eq.~\eqref{8der contra} we have
\beq
\mu^2_1 \mu_{1^\prime}^2 \mu^2_2 = ( a^4 + b^4 + 2a^2 b^2 )\mu^2_2 \sim M_P^6
\eeq
with either $a^2 \ll b^2 \sim M_P^2$ or $b^2 \ll a^2 \sim M_P^2$.

This means that
it is possible to have a strong hierarchy between the real and imaginary
parts of the roots $\mu_i$---with huge-mass parameters in the action.
Yet, we do not call it a working seesaw mechanism because the physical effects
of such a hierarchy does not imply a way out of the Planck suppression.
It only means that the largest among the real or the imaginary part
is going to dominate the phenomenology, and this quantity is on the order of
$M_P$. Indeed, if $b^2 \ll a^2 \sim M_P^2$, then to most practical purposes we
can assume that the pair $(\mu_1, \mu_{1^\prime})$ behave as degenerate
modes of mass $a \sim M_P$; while if $a^2 \ll b^2 \sim M_P^2$ they behave as
a degenerate tachyonic pair.

Moreover, this procedure proves that there is no natural
choice of massive parameters in the action which can provide
a simultaneous reduction of both real and imaginary parts of
the complex ``masses'' of the theory. In conclusion, there
is no seesaw-like mechanism efficiently working in the polynomial
higher-derivative gravity, even if the propagator contains complex poles.

In order to close this section it is instructive to explicitly work out the
aforementioned example of the six-derivative gravity~(\ref{Lag6orderGravity}). According to~\eqref{Def_masses}, the condition for
having complex poles in the propagator of the tensor modes reads
$\beta^2\ka^2 +16B < 0$. The ``masses'' $\mu_{2\pm}$ can now be written as
$\mu_{2\pm} = a_2 \mp i b_2$ so that
\beq
\label{a2_and_c2_def}
a_2^2
=
\frac{-\be + \sqrt{\frac{16\vert B \vert}{\ka^2}}}{4 \vert B \vert }\,,
\quad
b_2^2
=
\frac{\be + \sqrt{\frac{16\vert B \vert}{\ka^2}}}{4 \vert B \vert }\, .
\eeq

One can classify the possible situation as follows.
In case that $16|B|$ is only slightly larger than $\be^2\ka^2$
there is a strong hierarchy between real and imaginary parts and
the ``masses'' $\mu_{2\pm}$ tend to be approximately equal.
In fact, if $\be<0$ we get $a_2 \gg b_2$ and both excitations
behave almost like normal particles of the same mass, while $\be>0$
yields $a_2 \ll b_2$ and we have two tachyons. If
$16|B| \approx \be^2\ka^2 \sim M_P^{-2}$, then
\beq
\mu_{2\pm}^2 \approx  -\frac{8}{\beta\ka^2} \sim M_P^2.
\eeq
This figure can be reduced only by choosing a huge $|\beta|$ (and
simultaneously, in this case, a huge $|B|$).

On the other hand, if $\,16 |B| \gg \be^2 \ka^2$, there are
``masses'' with real and imaginary parts of the same order
of magnitude. This scenario is a truly complex one, however
since $a_2 \approx b_2$ one can work with
a single massive parameter, and therefore the only possibility
for a seesaw mechanism would be to reduce this quantity far
below the Planck mass. Notwithstanding, we have
\beq
a_2^2 \, \approx \, b_2^2
\, \approx \,
\sqrt{\frac{1}{\ka^2 |B|}} \,=\, \frac{M_P}{\sqrt{2|B|}}\,.
\eeq
Therefore, in order to $\, a_2^2 , b_2^2 \ll  M_P^2 \,$ hold, it is
necessary to impose $\,|B| \gg M_P^{-2}$. Thus, reducing the parameter $\be$
cannot diminish the effective mass and the only way of achieving
this is by increasing $|B|$ to unnatural values, that is, by applying the ``weak seesaw''
condition. It is interesting to notice that, in opposition to the real
poles weak seesaw (choosing a huge $|\beta|$, cf.~\eqref{8}), in the case of complex
poles it is a condition on $B$. A similar discussion applies to the complex scalar modes $\mu_{0\pm}$.

\section{On the physical consequences of gravitational seesaw}
\label{S3}

From the general perspective it is interesting to discuss what
could be the phenomenological consequences of the much lighter
massive ghost. Let us note that these and related subjects are
discussed in detail in the context of the general six-derivative
model in  the parallel paper \cite{Larger}. Here we present
just a brief extract of the results, which have relation to the
seesaw mechanism. Let us start from some obvious statements.

The presence of light excitations in the spectrum of the theory
would reduce the Planck suppression at both classical and
semi-classical cases, and bring the physical relevance of the
massive modes to the low energy domain. This would imply,
for example, modifications of Newton's inverse-square force
law \cite{Larger,Newton-high,Giacchini-poles} which is measured in
torsion-balance experiments~\cite{Kapner07,Giacchini14MGM}.
In the case of complex poles, for example, the corrections owed
to the higher-derivatives assume the form of oscillating terms, as
it was noticed in \cite{Larger,Modesto2016}. A stimulating discussion
on the perspective of detecting oscillations in the gravitational
potential can be found in \cite{Perivolaropoulos}. An important
theoretical feature of the higher derivative corrections to the
Newton gravitational law is that the relevant contributions
come from both tensor and scalar sectors of the theory
\cite{Stelle77,Newton-high}.

Another possibility of detecting signatures of higher derivatives
is the gravitational light bending. In this case the tensor and
scalar excitations play different roles. It was shown (see, e.g.,
\cite{Accioly15}) that the deflection of light in the four-derivative
gravity explicitly depends only on the tensor modes of metric perturbations.

It proves interesting to discuss this issue in full detail. Let us
start from the simplest case.  The linearized versions of general
relativity and the $R+R^2$ gravity model yield the same equations
of motion for photons \cite{BD0}.  On the other hand, the $R+R^2$
model is equivalent to the Brans-Dicke theory with a massive degree
of freedom. It is possible to show that the light bending alone cannot
distinguish between this type of metric-scalar gravity theories
and general relativity \cite{BD0,BD1,BD2,Bd3}. One can
find  in these references the discussion concerning the difference
with the massless Brans-Dicke theory \cite{Will-book}, where
the mass of the massive body (e.g., a star or galaxy) creating the
gravitational field must be renormalized.

One can easily understand the reason for the difference between
massive and massless Brans-Dicke theories in the framework
of the equivalent $R+R^2$ gravity model. The need of
rescaling the Newton's constant $G$ and/or of the measured
masses of astronomical bodies is that the Yukawa term in the
modified Newtonian potential~\cite{Stelle77}
\beq
V(r)
&=&
- \,\frac{GM}{r} \,\Big( 1\,+\,\frac13 \,e^{-\mu_0r} \Big)
\label{modNR2}
\eeq
becomes, in the massless or very light mass limit $\mu_0 \to 0$,
the same as the Newtonian term. Hence, effectively in the massless
limit one can replace $GM$ by $\,4GM/3$, at it occurs in the
massless Brans-Dicke.

The presence of a massive scalar mode
with $\mu_0>0$ also  affects the effective quantity $GM$, but
only at the range of distances below $1/\mu_0$. For the  massive
Brans-Dicke theories this was explicitly shown in
Ref.~\cite{Will-2012}.  For the properly chosen small $\mu_0$
this leads to a mismatch between an effective mass $M$ which
can be observed at the astrophysical and laboratory scales
\cite{PolStar2000} and may affect the predictions for the
deflection of light.
This result can be indeed generalized for the $R+R\Box R$ and
more general models with even more derivatives. At the same time,
in the case of higher derivative gravity models with six or more
derivatives such a light scalar with Compton wavelength at the
astronomical scale is out of the scope of the present work.
Therefore we assume much larger masses of the massive
gravitational modes.

The introduction of massive parameters in the tensor sector, on the
other hand, has a direct influence on the deflection of light. At
quantum level it yields scattering cross-sections that depend on the
energy of the photon, as it was shown, e.g., in
Refs.~\cite{Accioly15,Larger,Caldwell}.
This result, however, has little application for the deflection by
astronomical bodies~\cite{Larger}. From the classical perspective,
those tensor modes only play active role if the reciprocal of their
masses are comparable to the light ray impact parameter
(see~\cite{Accioly15} for a specific discussion on light bending
in the model with four derivatives, and~\cite{Larger}, for the one
with six derivatives and all possible scenarios for the massive
poles). In the case of the polynomial model with real poles in the
propagator, this would require a light ghost, which could be
provided by the (weak) seesaw mechanism. Nonetheless, light
deflection by the Sun can not yield better observational constraints
on such masses than torsion-balance experiments at laboratory
scales~\cite{Larger,Giacchini14MGM}.

To conclude the discussion on light deflection, if the seesaw
mechanism  works in at least one of the massive  sectors, the
effects of such light masses  are more likely to be  observed
in the  modified  Newtonian potential  than on the bending of
light in the Solar System.

Last but not least, a light ghost could affect the cosmology and
especially the stability of classical solutions with respect to tensor
perturbations. For the four-derivative models this issue was
discussed in \cite{GW-Stab,MSS,IT}. The absence of the effectively
working seesaw mechanism shows that the Planck protection
which was discussed in \cite{GW-Stab,GW-HD-MPLA} is
working the same way in the six-derivative (and higher
derivative) gravity.

\section{Conclusions and discussions}
\label{S5}

We have described a qualitatively new gravitational seesaw
mechanism which might be possible in the higher derivative
gravity models with the number of derivatives $\geq 6$. These
theories are characterized by a discrete spectrum of ``masses''
which may be real or complex.

If the dimensional parameters of the action have Planck order
of magnitude, they could combine, in principle, in such a way
that one of the masses is still on the order of $M_P$ while another
is many orders of magnitude smaller. As we have seen above,
such a strong gravitational seesaw is not possible. An essential
reduction of the mass of the lightest particle can be achieved only
by adjusting the four-derivative term in the action. Adding the
six-derivative terms does not modify the situation in this part.

A strong reduction of mass of the lightest tensor ghost
can be achieved by taking a huge value of the dimensionless
parameter $\be$, exactly like in the four-derivative gravity.
This situation is qualitatively similar to the one
in  the extra-dimensional theories. The difference is that here
the reduction of Planck suppression occurs due to the choice of
$\be$ and not because of the incomplete compactification
of some extra dimensions.

In the more realistic case of complex poles it is still possible
to extend the notion of a seesaw mechanism, which should now be
understood in the sense that the real massive parameters that
appear in the physical quantities of the theory are much below
the Planck scale. Nevertheless, we showed that this could only
be achieved by simultaneously reducing both real and imaginary
parts of one of poles of the propagator, which only happens with
small massive parameters in the action. Thus, only the weak seesaw
is possible even in the model with complex poles.
Further investigations on general higher-derivative models with complex
poles are carried out in~\cite{Giacchini-poles}.

Finally, we briefly mentioned some phenomenological aspects which could
be investigated in the presence of the (weak) seesaw mechanism;
namely, the modified Newtonian potential and the bending of light.
Further consideration on these issues can be found in~\cite{Larger},
whose main focus is on low-energy aspects of the general theory
with six derivatives.

Taking  into account the possibility to have a continuous mass
spectrum of the models such as \cite{Tseytlin-95}, the main
conclusions which we can draw at the moment are as follows:
\begin{itemize}
\item[i)] A strong gravitational seesaw does not work in the same
way like in neutrino Physics. One can not reduce the lightest
ghost mass by tuning the parameters of the higher derivative
action (\ref{superre}), except the dimensionless parameter $\be$.
\item[ii)] Since huge values of the dimensionless parameters
$\alpha$ and $\be$ cannot be completely ruled out theoretically,
it is important to derive the corresponding upper limits from
experimental and observational sides. In the present work we
made some step in this direction.
\item[iii)]
Our results indicate the importance of developing experimental
facilities for higher precision tests of the inverse-square force law,
since the detection of such an effect could provide a useful
information about higher derivatives in gravity.  In particular,
it looks relevant to explore the possibility of an oscillating
behaviour of the gravitational potential which is  typical for
the complex poles \cite{Larger}.
\item[iv)]
From the theoretical side, the generally negative result for the
gravitational seesaw mechanism is relevant for the alternative
approach for dealing with the ghost problem, suggested in
\cite{Simon-90}.
In this case the higher derivative terms and the corresponding
loop contributions are regarded as small corrections to the
Einstein-Hilbert action. This kind of approximation may work
only if all the ghost-like states belong to the far UV compared to
the scale of the gravitational phenomena. Keeping this in mind,
this is certainly an efficient ad-hoc scheme, not taking into account
the price to pay for it, which is that one can not deal with the
Planck-scale energies and that, for instance, the Starobinsky
model of inflation \cite{star,star83} must
be forbidden \cite{SimonStar}. Anyway, our results are very
relevant for this approach, since it can be extended to
the theories with more than four derivatives. As far as the
reduction of the effective gravitational ghosts masses does not
occur without unnatural choice of dimensionless parameters,
one can use the mentioned scheme in the more general kind
of theories.

\end{itemize}

\section*{Acknowledgements}
B.L.G. and I.Sh. are very grateful to Prof. Starobinsky for critical 
observations and useful discussions, which helped us to improve 
the paper. 
A.A. is very indebted to CNPq and FAPERJ, while B.L.G. is thankful  
to CNPq for supporting his Ph.D. project.
I.Sh. is grateful to CNPq, FAPEMIG and ICTP for partial support
of his work. Part of it was done during his visit to ICTP and it is
a pleasure to thank the high-energy group for warm hospitality.


\end{document}